# Topological lasers generating and multiplexing topological light

B. Bahari, L.-Y. Hsu, S. H. Pan, D. Preece, A. Ndao, A. El Amili, Y. Fainman, and B. Kanté[*]

*Department of Electrical and Computer Engineering, University of California San Diego, La Jolla, CA 92093-0407, USA*

[*]bkante@ucsd.edu

**Abstract-** Vortices are topologically stable singularities at the center of a swirl of energy. Optical vortices are conventionally formed using diffractive optics or by bespoke optical elements. We report room temperature integrated lasers directly generating and multiplexing coherent beams carrying arbitrarily large orbital angular momenta (OAM). The OAM beams are created using two-dimensional topological-rings formed by circular boundaries between topologically distinct photonic materials that naturally radiate vortices in the third dimension. We also demonstrate the planar multiplexing of OAM beams using concentric lasers. Our experimental demonstration reveals a subtle connection between topological matter and topological light and provides opportunities in microscopy, metrology, high-capacity communications, and quantum information processing.

Vortices are waves that possess a phase singularity and a rotational energy flow around a singular point or line [1-2]. They are ubiquitous in physics and have been observed in systems ranging from atmospheric scale tornadoes to atomic scale Bose-Einstein condensates [3]. Optical vortices are beams carrying orbital angular momentum (OAM). They possess a donut-shaped intensity profile and a helical phase structure with an azimuthal phase dependence of the form exp(i$\ell$θ), where $\ell$, an integer called topological charge, indicates the number of times and the direction in which the beam twists about its axis in one wavelength [4]. Beams carrying OAM promise revolutionary applications from metrology and particles manipulation, to enhanced resolution optical imaging [5-8]. Helical wavefronts are usually formed using bulky devices such as spiral phase plates [9-10], cylindrical lens converter [11], Q-Plates [12-13], holograms [14], and spatial light modulators [15-17]. These conventional techniques, including recently reported metasurfaces [18-22], require an external input beam originating from a separate light source. Latterly, an OAM laser with a topological charge of one, based on PT-symmetry, was reported [23]. It emits a beam with an angular momentum of $\ell$=M-N, where M is the azimuthal resonant order of a ring and N is the total number of grating elements used to scatter light out. The formation of pure OAM states requires M and N to be close, thus limiting such techniques to small topological charges. Albeit the theoretically unbounded OAM basis (limited by the size of the aperture) makes it appealing in areas such as high-capacity communication, quantum cryptography, or increased resolution in particles motion detection, sources capable of generating arbitrarily large OAM have been challenging [24-27]. Furthermore, the possibility to directly multiplex different lasers emitting OAM in an integrated device has remained elusive.

Here we report integrated lasers generating arbitrarily large topological charges that can be multiplexed in a planar device. The OAM beams are created using two-dimensional topological-rings formed by circular boundaries between topologically distinct photonic crystals (PhCs) that naturally radiate vortices in the third dimension under optical pumping at room temperature. We also demonstrate the planar multiplexing of OAM beams using concentric lasers. Our experimental demonstration reveals a subtle connection between topological matter and topological light and provides opportunities in microscopy, metrology, high-capacity communications, and quantum information processing.

The large orbital angular momentum and its multiplexing platform are presented in Fig. 1a. It is made of circular boundaries between two topologically distinct PhCs forming two-dimensional cavities that we call topological-rings. The dissimilar topologies of the photonic structures inside and outside a ring ensures the existence of one-way circulating edge states. In conventional ring resonators, whispering gallery modes (WGMs) are excited in pairs (clock-wise and counter clock-wise), resulting in zero net angular momentum. Topological-rings made of 2.5D photonic structures are leaky-wave emitters and thus radiate in the third dimension (i.e. out-of-plane). As such, they do not necessitate additional scattering elements to extract light from the cavities. The propagation phase offsets at different points of the traveling wave around the leaky-ring results in the formation of OAM beams in which the topological charge is equal to the azimuthal resonant order of the ring (see SI). The topological charge can thus be made arbitrarily large with the radii of rings. By alternating concentric circular boundaries between two PhCs of distinct topologies, an arbitrary number of orthogonal OAM beams of alternating chiralities can be multiplexed in a planar manner using a single aperture. This constitutes the first approach directly multiplexing OAM lasers.

Photonic crystal one (PhC1) is formed by a four-armed star-shaped unit-cell and has a non-trivial band gap with a non-zero Chern number, |ΔC|=1. Photonic crystal two (PhC2) has a triangular lattice with cylindrical air hole unit-cell and a trivial band gap with a zero Chern number. The topologically non-trivial structure is obtained by bonding InGaAsP multiple quantum wells (MQWs) on yttrium iron garnet (YIG). The YIG substrate is used to break time-reversal symmetry in the system under a static external magnetic field (MF) that opens the non-trivial gap of PhC1. Circular boundaries between PhC1 and PhC2 are leaky-wave topological-rings (ring 1, ring 2, and ring 3 illustrated in Fig. 1) supporting one-way edge states (see SI). The number of edge modes inside the ring is equal to the difference between the Chern numbers of PhC1 and PhC2 [28]. Therefore, with a Chern number difference of one, only one unidirectional edge mode can propagate at the boundary of the two PhCs, making the cavity single mode, an important attribute for stable laser operation [29-32]. Figure 1b presents the theoretical prediction of the topological charge $\ell$ of light as a function of the normalized radius of the topological-ring $a/\lambda_{eff}$, where, $a$ is the radius of the ring and $\lambda_{eff}$ the wavelength of the edge mode (see SI). Theoretical far-field patterns of topological-rings of various charges $\ell_1$, $\ell_2$, $\ell_3$, and $\ell_n$ are shown as insets and correspond, as expected, to doughnut-shaped beams of increasing radii. The chirality of the edge states, i.e. the sign of the topological charge, can be controlled by simply reversing the static external MF (see SI).

Figure 2 presents scanning electron micrographs of topological-ring OAM lasers. The structures are fabricated by electron beam lithography followed by dry etching and then bonded on the YIG substrate using a thin layer of polymethyl methacrylate. The InP substrate, on which InGaAsP is epitaxially grown, is subsequently removed by wet etching using hydrochloric acid (see SI). The multiplexed topological-rings, realized in one-step lithography, consist in the planar integration of ring 1, ring 2, and ring 3 in a single aperture. An arbitrary number of rings can thus be multiplexed using this principle. It is worth noting that there is a minimum number of unit-cells needed to create a band gap which in turn imposes a minimum radius that the topological ring should have (see SI). A finite aperture would also put a bound on the maximum topological charge. The non-trivial (PhC1) and the trivial (PhC2) photonic crystals can be seen in Figs.2. The circular boundary encloses PhC1 in ring 1 (Fig. 2a), PhC2 in ring 2 (Fig. 2b), and PhC1 in ring 3 (Fig. 2c) resulting in edge states propagating clockwise in rings 1 and 3, and counter-clockwise in ring 2. The topological charges of the fabricated rings are $|\ell_1|$=100, $|\ell_2|$=156, and $|\ell_3|$=276 and they are naturally orthogonal states.

The lasers are characterized by optically pumping the entire surface of the crystals in presence of a static external MF of **B**=+B$_0$**e**$_z$ (normal to the PhCs plane) with B$_0$ = 100 Oe that saturates the YIG material. An edge mode with a frequency within the band gap of the two PhCs is excited and confined at the interface of PhC1 and PhC2. The absence of band gap in the direction perpendicular to the structure plane for edge modes above light-line results in natural out-of-plane scattering of the one-way traveling wave around the cavity. The OAM beams originate from the interference of those leaky waves. This is a fundamental difference between our topological-rings and conventional ring-resonators using WGM [23,33], as the latter requires a large number of gratings to create out-of-plane scattering and maintain mode purity. In our system, the generated topological charge is exactly the azimuthal number of the WGM, naturally enabling large charges. The micro-photoluminescence setup pumps the sample from the top with a pulsed laser ($\lambda_{pump}$ = 1064 nm, T = 6 ns pulse at a repetition rate of 290 kHz) using an objective lens. The emission of the devices is collected using the same objective lens and directed to a monochromator for spectrum analysis or to an infrared camera for imaging.

Figure 3a-c shows the photoluminescence spectrum of optically pumped topological-ring 2 in presence of $B_0$ in three different regimes. The amplification and selection of a single mode at the wavelength of λ=1575 nm are observed with increasing pump power. Similar results were observed for other topological rings. Figure 3d presents the log-scale output power as a function of the pump power density (light-light curve) and shows a characteristic lasing behavior. To further investigate the coherent character and lasing characteristic of the cavity, we measured the second-order intensity correlation function of its emission, $g^2(\tau) = \frac{\langle I(t)I(t+\tau)\rangle}{\langle I(t)\rangle^2}$, using a Hanbury Brown-Twiss interferometer (see SI). $\langle I(t)\rangle$ represents the expectation value of the intensity at time $t$. Figure 3e shows the zero-delay of the normalized second-order intensity correlation function, $g^2(0)$, and the three different regimes of spontaneous emission (SpE), amplified spontaneous emission (ASE), and stimulated emission (StE) are evidenced. We observe the suppression of the photon bunching peak (visible in the ASE regime) in the StE regime, i.e., lasing action (see SI). Furthermore, the full width at half maximum (FWHM) of the zero-delay $g^2(\tau)$ pulse shrinks in the SpE regime, reaches a minimum in the ASE regime, and broadens in the StE regime (Fig. 3f). Such variations in the $g^2(\tau)$ pulse width are related to a nonlinear effect called delay threshold phenomenon or dynamical hysteresis, which occurs in a laser only when the peak intensity of a pump pulse is larger than the threshold intensity [34]. The distinct $g^2(\tau)$ width behaviors in the SpE and StE regimes signify that the suppression of the photon bunching peak (Fig. 3e) at high pump intensity indeed originates from lasing instead of SpE. It is worth noting that the unity $g^2(0)$ in the SpE regime is due to the limited time resolution of the detection system (see SI) [34]. Similar results are observed for other topological-rings.

Far-field intensity patterns of the beams generated by topological-rings 1, 2, 3 and their multiplexing, under a pump power density of ρ=1.03 µW/µm², i.e. in the stimulated emission regime, are presented in Figs. 4a-d. This experiment excites all three rings using one optical pump, but, the lasers should in principle be driven independently for communication applications. They clearly exhibit ring-shape profiles. Figure 4e-f present far-field interference patterns of the emission from topological-ring 2 of charge $\ell_2$ with the opposite topological charge $-\ell_2$, obtained theoretically (Fig. 4e), and experimentally (Fig. 4f). The opposite charge is achieved experimentally by reflecting the beam on a mirror. The interference patterns evidence fringes characteristic of beams carrying orbital angular momenta. The fringes are clearly observed in the insets of Fig. 4e-f and the purity of generated modes is discussed in SI. Topological-ring 2 is designed for a charge of $|\ell_2|$=156 and the total number of measured fringes is 312. The interference patterns of other rings are measured in a similar manner (see SI). These results demonstrate the successful generation of coherent OAM beams of large charges.

We demonstrated experimentally topological ring-resonators emitting coherent beams carrying orbital angular momenta of arbitrarily large topological charges. Topological-rings are formed by circular boundaries between topologically distinct photonic structures and they constitute leaky-wave sources naturally radiating orthogonal orbital angular momenta states. Those states are multiplexed by integrating concentric rings emitting waves of controllable chirality. We have also demonstrated the coherent property of the proposed lasers by measuring their second-order intensity correlation. These results demonstrate that topological matter can be used to uniquely generate topological light and open the way to integrated lasers emitting on demand far-field patterns. Such sources will find applications in classical and quantum optics for communications, sensing, or imaging.

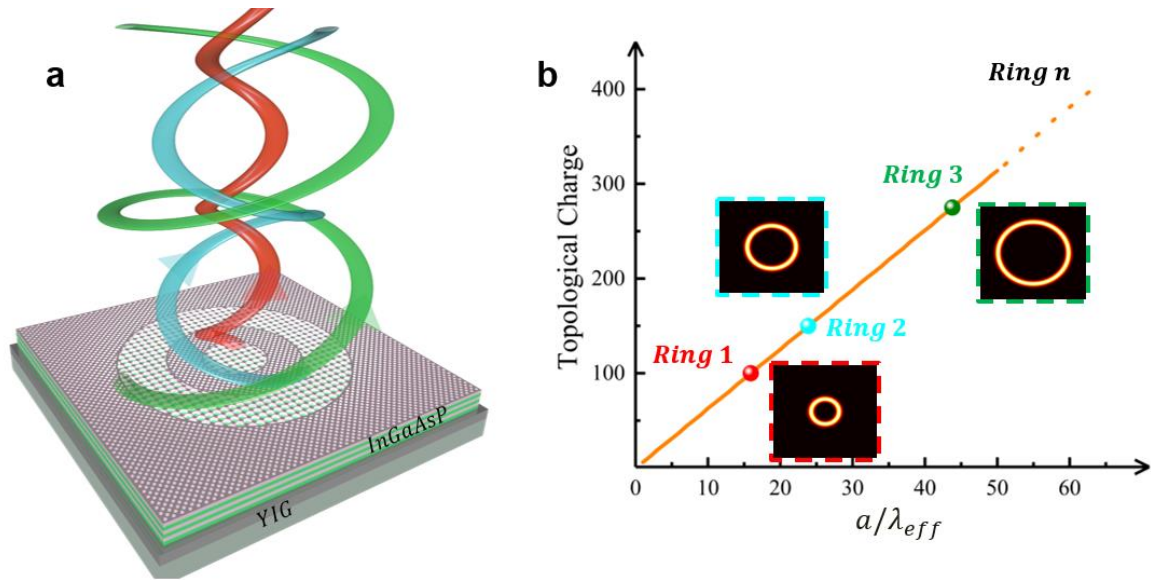

**Figure 1 | Multiplexing of orbital angular momenta topological-rings. a,** Schematic of concentric topological-rings formed by circular boundaries between topologically distinct photonic band gap materials. The square lattice photonic crystal (PhC1) has a four-armed star-shaped unit-cell and a non-trivial band gap with a non-zero Chern number, $|\Delta C|=1$. The triangular lattice photonic crystal (PhC2) has a cylindrical air hole unit-cell and a trivial band gap with a zero Chern number. The non-trivial topology is obtained by bonding InGaAsP multiple quantum wells (MQWs) on yttrium iron garnet (YIG) under a static magnetic field normal to the PhCs surface **B**=$B_0$**e**$_z$ . Circular boundaries between PhC1 and PhC2 are a leaky-wave topological-rings (ring 1 to ring n) supporting one-way edge states (see SI). The propagation phase offsets at different points of the traveling wave around the leaky-ring results in the formation of orbital angular momenta (OAM) beams in which the topological charge is equal to the azimuthal resonant order of the ring. The topological charge can thus be made arbitrarily large with the radii of rings. By alternating circular boundaries between PhC1 and PhC2, an arbitrary number of orthogonal OAM beams of alternating signs can be multiplexed in a planar manner using a single aperture. **b,** Theoretical prediction of the topological charge $\ell$ as a function of the normalized radius of the rings $a/\lambda_{\text{eff}}$, where $a$ is the radius of the ring and $\lambda_{\text{eff}}$ the wavelength of the one-way edge mode. Insets show the theoretical far-field patterns for topological-rings of various charges with doughnut-shaped beam patterns. The chirality of the edge states can be controlled by the static external magnetic field.

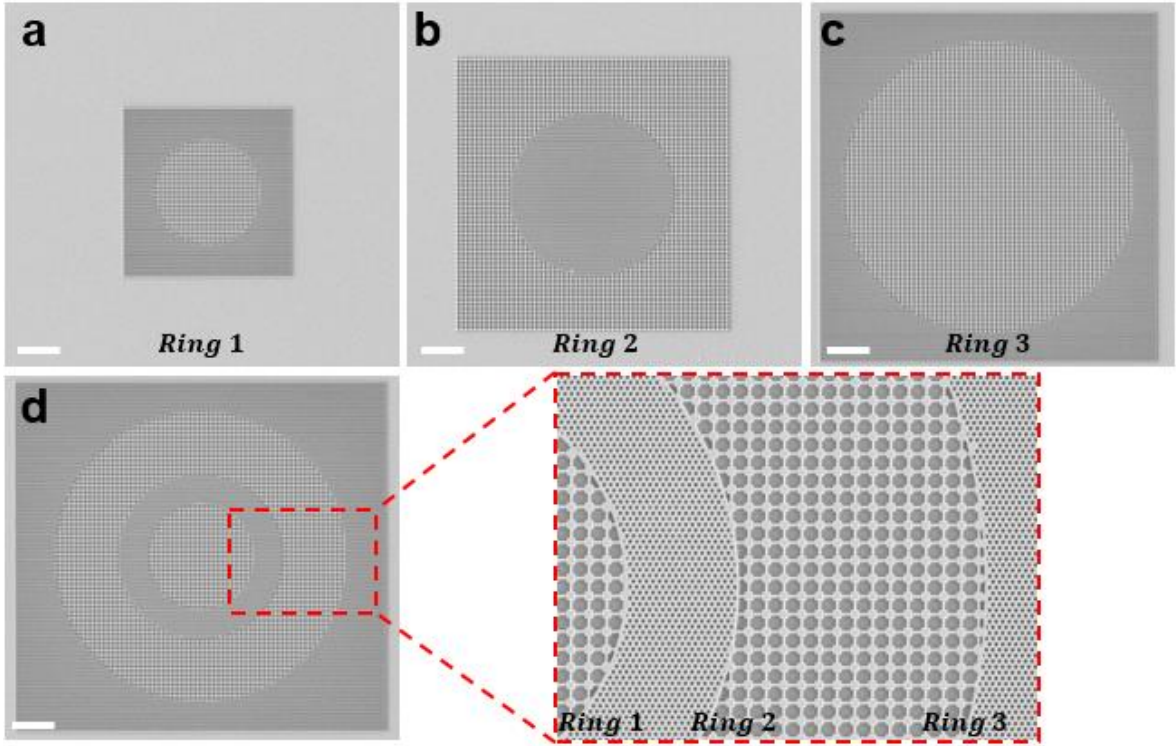

**Figure 2 | Scanning electron micrographs of topological-ring OAM lasers.** Top view scanning electron micrographs of topological-rings formed by the non-trivial (PhC1) and the trivial (PhC2) photonic crystals. The circular boundary encloses PhC1 in ring 1 (**a**), PhC2 in ring 2 (**b**), and PhC1 in ring 3 (**c**) resulting in edge states propagating clockwise in ring 1 and 3, and counter-clockwise in ring 2. The topological charges of the fabricated rings are $|\ell_1|=100$, $|\ell_2|=156$, and $|\ell_3|=276$.

The chirality of the vortices (sign of the charge) can be switched by reversing the propagation direction of the edge mode with the external magnetic field. The structure is fabricated on InGaAsP by electron beam lithography followed by dry etching and then bonded on a YIG substrate with a thin layer of polymethyl methacrylate. The InP substrate, on which InGaAsP is epitaxially grown, is subsequently removed by wet etching using hydrochloric acid (see SI). **d,** Multiplexed integrated rings corresponding to the planar superposition of ring 1, ring 2, and ring 3. An arbitrary number of rings can be multiplexed using this principle. The scale bars represent 10 μm and the unit-cells of PhC1 and PhC2 have a periodicity of p = 1084 nm and p/3 respectively.

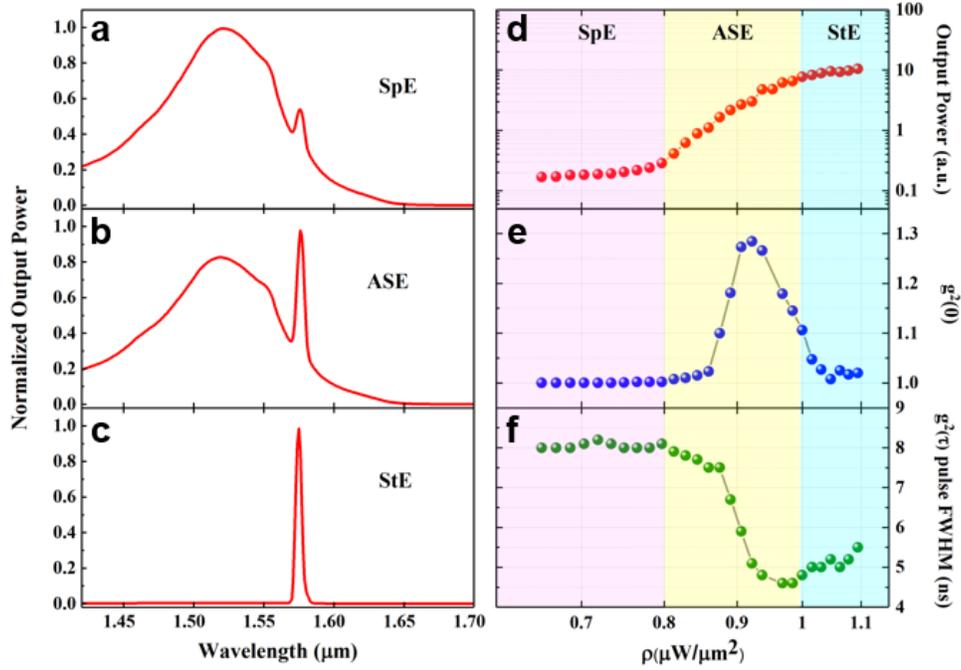

**Figure 3 | Experimental characterization of the topological OAM sources.** Photoluminescence spectrum of an optically pumped topological-ring (ring 2) in presence of an external magnetic field of $B_0$ =100 Oe in the (**a**) spontaneous emission (SpE), (**b**) amplified spontaneous emission (ASE), and (**c**) stimulated emission (StE) regimes. The surface of the laser is uniformly pumped with a high-power pump laser at 1064 nm. The amplification and selection of a single mode are observed when pump power is increased. **d,** Output power as a function of the pump power (light-light curve). **e,** Second-order intensity correlation at the zero-delay, $g^2(0)$ and, (**f**) its full width at half maximum (FWHM). Clearly, the photon bunching peak (in the ASE regime of $g^2(0)$) is suppressed in the StE regime, and, the FWHM of the $g^2(\tau)$ shrinks, reaches a minimum in the ASE regime and broadens in the StE regime, unambiguously demonstrating lasing action of the topological-ring (see SI).

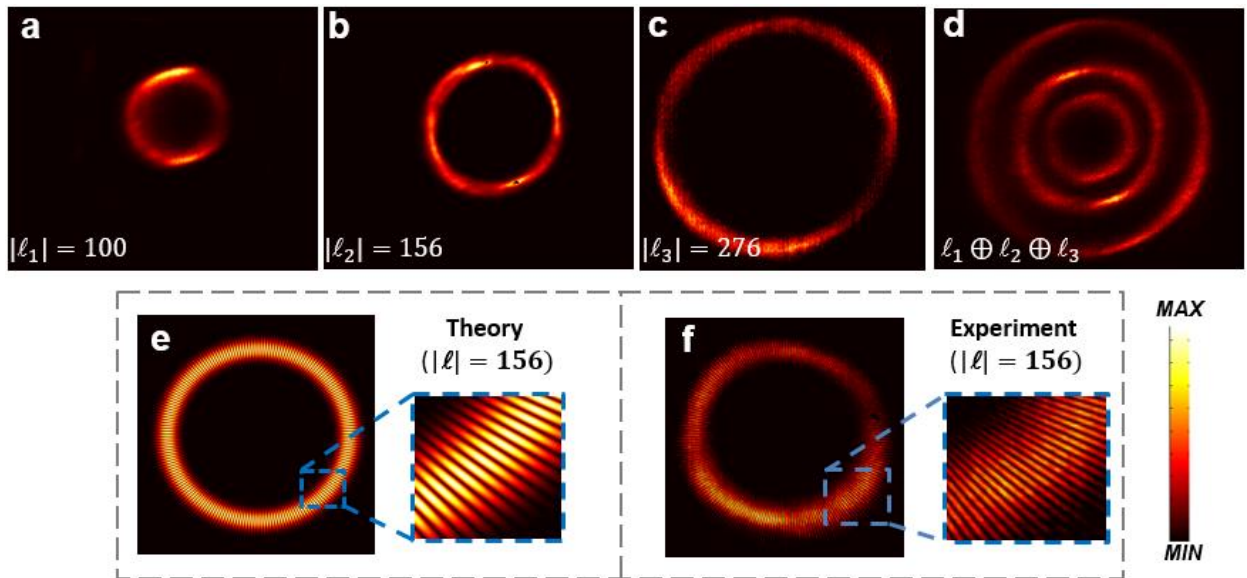

**Figure 4 | Far-field intensity and interference patterns of the OAM beams.** Real-space camera images of the optically pumped OAM lasers. Ring 1, 2, and 3 present clear ring-shape patterns in **a**, **b**, and **c** respectively. **d**, Multiplexed lasers with three concentric beams of charge $\ell_1$, $\ell_2$, and $\ell_3$. Far-field interference pattern of the emission from ring 2 of charge $\ell_2$ with the opposite topological charge $-\ell_2$ obtained theoretically (**e**), and experimentally (**f**). The opposite charge is achieved experimentally by reflecting the beam on a mirror. The interference patterns evidence fringes characteristic of beams carrying orbital angular momenta. The fringes are clearly observable in the insets. Topological-ring 2 is designed for a topological charge of $|\ell_2|=156$ and the total number of the fringes is 312. The interference patterns of rings 1 and 3, measured in a similar manner, are presented in supplementary information (see SI).